\documentclass[fleqn,twoside]{article}
\usepackage{espcrc2}

\usepackage{graphicx}
\usepackage[figuresright]{rotating}

\title{Collective transport and optical absorption near the 
stripe criticality}
\author{C. Di Castro\address{Dipartimento di Fisica, Universit\`a di 
Roma ``La Sapienza'',
and INFM Center for Statistical Mechanics and Complexity,
Piazzale Aldo Moro 2, I-00185 Roma, Italy}
\thanks{We acknowledge financial support from MIUR COFIN 2001, prot.
2001023848, and INFM/G PA-G0-4.}, S. Caprara\addressmark, 
and M. Grilli\addressmark}

\begin{document}

\begin{abstract}
Within the stripe quantum critical point scenario for high $T_c$
superconductors, we point out the possible direct contribution
of charge collective fluctuations to the optical absorption and to
the d.c. resistivity. 

\end{abstract}

\maketitle

There are by now several experimental evidences 
that the peculiar 
properties of the cuprates
are controlled by a Quantum Critical Point (QCP), located near
(actually slightly above) the 
optimal doping $\delta=\delta_{opt}$ \cite{TALLON}.
Already several years ago some of us proposed \cite{CDG}
that the pseudogap region of the underdoped cuprates
is characterized by the pronounced
tendency to form (local) spatial ordering (the so-called ``stripe'') phase
below some Charge-Ordering (CO) temperature $T_{CO}(\delta)$.
This last temperature
can therefore be related to the temperature $T^*(\delta)$
at which the pseudogap forms.
As a schematic model, we adopted the Hubbard-Holstein
model in the presence of long-range Coulombic forces.
With reasonable parameters for the cuprates, for
infinite Hubbard repulsion  and for moderate
electron-phonon coupling there occurs a mean-field
zero-temperature charge-density wave instability
with a finite modulating  wavevector ${\bf q}_c$ and 
below a rather large critical doping $(\delta_c^{(0)}
\sim 0.2-0.3)$ (see Ref. \cite{andergassen}
and references therein). This mean-field 
CO instability extends at finite temperatures below a doping dependent
temperature $T_{CO}^{(0)}(\delta)$, which rapidly increases towards
electronic energy scales upon underdoping.
We then considered
how $(\delta_c^{(0)}$ and $T_{CO}^{(0)}(\delta)$ are modified
by dynamical charge fluctuations. The critical doping $\delta_c$ is reduced
towards optimal doping and $T_{CO}^{(0)}(\delta)$ is reduced to
lower values $T_{CO}(\delta)$ 
 (typically a few hundreds of Kelvin and vanishing at the CO-QCP at
$\delta_c$). Owing to
the dynamical character of the charge fluctuations, $T_{CO}$ (and
the related $T^*$) were shown to display a marked dependence on the
typical time-resolution of the experimental probes (taken as an infrared
cutoff of the theory) and peculiar
isotopic dependencies \cite{andergassen}.

%\section{Optical conductivity and transport}
Recently  \cite{CDFG} we considered the possibility that
nearly critical charge fluctuations could provide an independent
absorption channel in parallel with the usual Drude quasiparticle 
(QP) absorption.
We explicitly calculated the current-current response function related
to the excitation of two collective modes as represented in Fig. 1(a)
in the form of an Aslamazov-Larkin-type of diagram. 
An additional peak appears at the finite frequency
needed to excite two charge collective modes
at ${\bf q} \sim \pm {\bf q}_c$.
This allows for a simple interpretation of low energy
absorption peaks observed in several overdoped compounds  \cite{OVD}.
These broad absorption peaks occur at low energy (below 300$cm^{-1}$),
and soften upon decreasing $T$,
thereby suggesting a collective character of the absorbing excitations.
Our theoretical analysis showed that the critical character
of the charge excitations entails a scaling form for the finite-frequency
absorption. This expectation is well verified experimentally, see Fig. 3
of Ref. \cite{CDFG},
and is accompanied by the temperature behavior of the absorption peak
frequency (see inset of Fig. 3 in Ref. \cite{CDFG}),
which is proportional to the expected ``mass'' 
$m(T,\delta)=\alpha T+ \beta (\delta - \delta_c)+O(T^2)$
for a critical mode at the crossover from a quantum critical 
($m\propto T$) to a quantum disordered region
($m\propto (\delta-\delta_c)$).
This point of view corresponds to the idea that
optimally and overdoped cuprates, when approaching the pseudogap region,
are a ``melting pot'' of fermionic
QP and collective critical excitations, both contributing
to the optical properties.  We here adopt this point of view
considering the charge CM contribution to transport.
According to the diagram of Fig. 1(a), 
the result for the resistivity due to collective transport is
\begin{equation}
\rho_{CM}(T) \propto T\mu^2(T,\delta)\left[ a+b\mu(T,\delta) \right]
\end{equation}
Here $a=1.1$ and $b=0.4$ are numerical coefficient, 
while $\mu(T,\delta)\equiv m(T,\delta)/T$.
In Fig. 1(b) we plot three typical curves refering to the overdoped,
critical, and underdoped regimes.
%%%%%%%%%%%%%%%%%%%%%%%%%%%%%%%%%%%%%%%%%%%%%%%%%%%%%%%%%%%%%%%%
\begin{figure}
\includegraphics[angle=-90,scale=0.5]{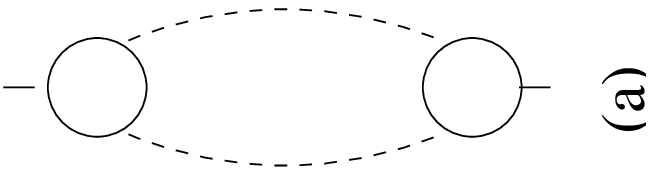}
\hspace {1. truecm}
\includegraphics[angle=-90,scale=0.2]{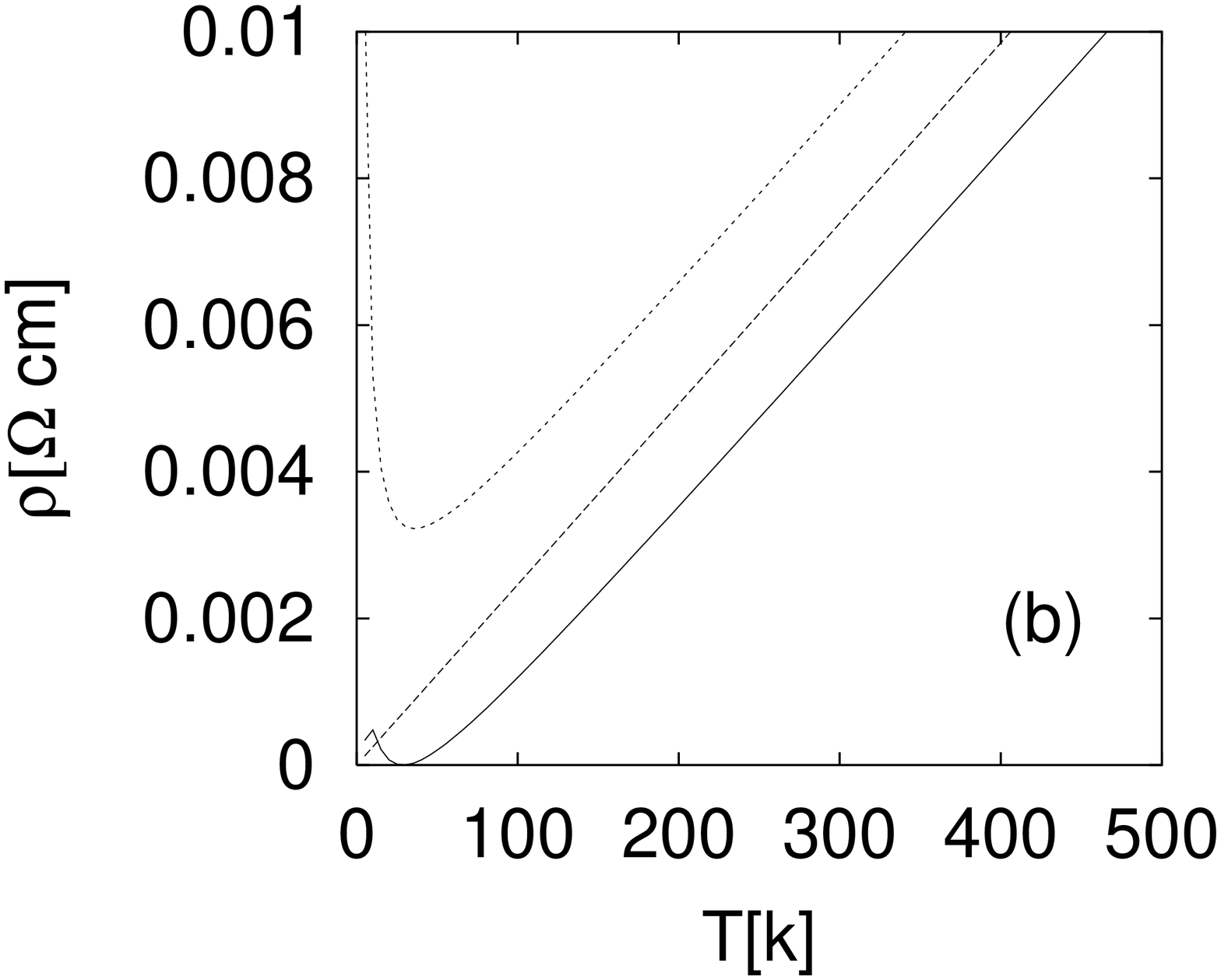}
%{\psfig{figure=rho.ps,width=5.5cm,angle=-90}}
\caption{(a) CM contribution to the current-current response function. 
(b) CM resistivity at $\delta=0.22$ (dotted), 
$\delta=\delta_c=0.19$ (dashed) and $\delta=0.16$ (solid).
The parameters of the model calculations are adjusted to yield
typical values for the cuprates.}
\end{figure}
%%%%%%%%%%%%%%%%%%%%%%%%%%%%%%%%%%%%%%%%%%%%%%%%%%%%%%%%%%%%%%%%
The most apparent feature is that $\rho_{CM}(T)$ is linear 
in the quantum critical region
down to $T=0$ for $\delta=0.19$ and stops at finite temperature
in the other two cases.
This is suggestive of the linear behavior of the
resistivity in nearly optimally doped cuprates. On the other hand,
on the overdoped region, when the temperature decreases below
the finite mass $\beta(\delta-\delta_c)>0$, the QP's in the fermionic
loops of Fig. 1(a) no longer can emit and reabsorb these
``massive'' CM's and this transport 
channel is suppressed. At this stage $\rho_{CM}(T)$ diverges and the
transport from QP's will eventually take over.
On the opposite doping side, $\rho_{CM}(T)$ seems to vanish
at a finite temperature of $T_{CO}(\delta)$ (near $T^*(\delta)$),
where the mass of the CM's also vanishes. Of course, 
our perturbative  theory can not be pushed down to this regime, where
other contributions should be taken into account.

We point out that several issues have to be addressed before definite 
conclusions can be drawn. First of all, here we fully neglected the 
parallel transport channel arising from QP's. 
This conventional transport channel would lead to
a $T^2$ resistivity behavior, and whenever the CM resistance is not
much lower than the QP one, a curvature in $\rho(T)$ will surely
arise in the total resistivity. On general grounds we expect that
moving towards the overdoped region, the QP transport will
eventually dominate over the collective one.
Secondly, in this preliminary analysis of $\rho_{CM}$,
we only considered the {\it leading} (i.e., critical) temperature and doping
dependencies. For instance we neglected the temperature and doping 
dependence of the fermionic loops of the diagram in Fig. 1(a) as well
as the doping dependence of the current vertices, which should be
present in a strongly correlated system. While
this is acceptable to identify some overall tendencies (like $\rho_{CM}
\propto T$),
these dependencies must be considered to (hopefully) account for
non-universal aspects like the doping dependence of
the slope of the linear resistivities and the downturn in $\rho(T)$
in underdoped cuprates.
All these issues are presently under investigation.

% \appendix

% \section{}
% \label{}

\end{document}